# Reaction front occurrence on imperfection profiles during oxygen chemical diffusion in oxides. II. Kinetic analysis

M. Sinder*, Z. Burshtein, and J. Pelleg

*Materials Engineering Department, Ben-Gurion University of the Negev, P.O.*

*Box 653, Beer-Sheva 84105, Israel*

Abstract

We present a theoretical study of the impact of oxygen diffusion in oxide crystals on metal dopants ionic state and the conduction type under dynamic changes. The slow changes under thermal equilibrium are provided in a companion, Part I paper. Oxygen vacancy formation acting as a shallow, double-electronic donor, is assumed to result from the crystal exposure to a low ambient oxygen pressure. Critical transitions from n- to a p-type at an oxygen partial pressure $P_i$, and in ionization state of the metal dopant at an oxygen partial pressure $P_M$, are usually not simultaneous, and depend on the different reaction constants. Invoking ambipolar diffusion for all participating species, it is shown that oxygen vacancy diffusion exhibits well defined characteristics in specific regions of the vacancy concentration related to the said critical pressures. Particularly, the oxygen chemical diffusivity is practically constant between $P_i$ and $P_M$, and increases sharply for pressures $P << P_i, P_M$ and $P >> P_i, P_M$. Prominent reaction fronts occur under specific conditions in the two latter regions.

**Keywords**:  Oxides; Diffusion; Defects; Color centers; Transport properties.

*Corresponding author. email address: micha@bgu.ac.il

# I. INTRODUCTION

This paper is a companion continuation of a paper published in this issue [1]. Doubly ionized oxygen vacancy diffusion through the oxide crystal medium induces valence transformations in resident dopant metal ions. The present paper deals with kinetics of processes related to sudden gross changes in the ambient partial oxygen pressure. Recently, it has been demonstrated [2-8] that improved insight of the process is obtained by utilizing the reaction rate, reaction zone, and reaction front concepts. In the present paper, the above theoretical techniques are used for a comprehensive analysis of the oxygen chemical diffusion in a metal-doped oxide crystal. Invoking ambipolar diffusion for all participating species, it is shown that oxygen vacancy diffusion exhibits well defined characteristics in specific regions of the vacancy concentration. Particularly, the oxygen chemical diffusivity is practically constant between $P_i$ and $P_M$, and increases sharply for pressures $P \ll P_i, P_M$ and $P \gg P_i, P_M$ (for detailed definition of $P_i$ and $P_M$, see Part I). We utilized Seeger's parametric solution of the diffusion equations [9] for the two latter regions. The use of Seeger's approach to both regions provides an essential improvement in the present article relative to Ref. [8], where it was used only for the $P \ll P_i, P_M$ region. Prominent reaction fronts occur there under specific conditions.

# II. KINETIC ANALYSIS

When the crystal is subjected to a change in the ambient oxygen pressure, oxygen vacancy diffusion is activated, inducing concentration changes of the other species (metal ions, free electrons and free holes) until a new equilibrium is established. In the following we consider the dynamics of these changes. Under local electro-neutrality and quasi-equilibrium of all chemical reactions, the oxygen chemical diffusivity $D^\delta$ should be used in the diffusion equations. Following Maier [10, 11],

$$D^{\delta} = \frac{1}{\dfrac{1}{4[V_O]D_V} + \dfrac{1}{[e^-]D_e + [h^+]D_h}} \times \left[ \frac{1}{4[V_O]} + \frac{1}{(1+R_e)[e^-]+[h^+]} \right] \ , \qquad (1)$$

where $D_V$, $D_e$, and $D_h$ are the diffusion coefficients of the free vacancies, free electrons and free holes, respectively, and $R_e \equiv [M]K_3 / (K_3 + [e^-])^2$. One usually assumes $D_e, D_h \gg D_V$. The vacancy diffusion equation then is [11]

$$\frac{\partial V(x,t)}{\partial t} = \frac{\partial}{\partial x} \left[ D^{\delta}\big(V(x,t)\big) \frac{\partial V(x,t)}{\partial x} \right] \ , \qquad (2)$$

where $V \equiv [V_O]$. Once Eq. (2) is solved, the other species concentrations are obtained using Eq. (9) Part I.

Tables 1 and 2 summarize the approximate expressions for $D^{\delta}$ using Eq. (9) in Part I, when $\sqrt{K_2} \gg K_3$, and when $K_3 \gg \sqrt{K_2}$, respectively, for different regions of the ambient partial oxygen pressure. Also specified are the reaction equilibrium states for each region. The oxygen chemical diffusivity is constant between $P_i$ and $P_M$; beyond, it always depends on the pressure $P$, hence on the vacancy concentration $V$.

TABLE 1: Approximate oxygen chemical diffusivity and reaction equilibrium states for different partial ambient oxygen pressure regions when $P_i \ll P_M$ ($\sqrt{K_2} \gg K_3$).

| $P$ | $P \ll P_i$ [a] | $P_i \ll P \ll P_M$ | $P_M \ll P$ [b] |
|---|---|---|---|
| $D^{\delta}$ | $D_e \dfrac{K_3[M]}{([M]-2V)^2}$ $= \dfrac{2D_e K_1 [O^{2-}][M]}{K_3} \cdot P^{-1/2}$ | $D_h \dfrac{K_2}{[M]K_3}$ | $D_h \dfrac{3}{8} \dfrac{K_2[M]}{K_3 V^2}$ $= \dfrac{3D_h K_2 K_3^{1/3}}{2^{5/3} K_1^{2/3}[O^{2-}]^{2/3}[M]^{1/3}} P^{1/3}$ |
| Reaction equilibrium State | $M^{p+} + e^- \overset{\longleftarrow}{\phantom{xx}} M^{(p-1)+}$ | $M^{p+} \overset{\longleftarrow}{\phantom{xx}} M^{(p-1)+} + h^+$ | $M^{p+} \overset{\longrightarrow}{\phantom{xx}} M^{(p-1)+} + h^+$ |

[a] Provided $P \gg 4[O^{2-}]^2 K_1^2 / K_3^2[M]^4$; $(D_e/D_V)^4[O^{2-}]^2 K_1^2 / 4[M]^6$.

[b] Provided $P \ll 32(D_V/D_h)^3[M][O^{2-}]^2 K_1^2 / (K_2^3 K_3)$.

TABLE 2: Approximate oxygen chemical diffusivity and reaction equilibrium states for different partial ambient oxygen pressure regions when $P_M \ll P_i$ ( $K_3 \gg \sqrt{K_2}$ ).

| P | $P \ll P_M$ [a] | $P_M \ll P \ll P_i$ | $P_i \ll P$ [b] |
|---|---|---|---|
| $D^\delta$ | $D_e \dfrac{K_3[M]}{([M]-2V)^2}$ $= \dfrac{2D_e K_1 [O^{2-}][M]}{K_3} \cdot P^{-1/2}$ | $\dfrac{3}{2} D_e \dfrac{K_3}{[M]}$ | $D_h \dfrac{3}{8} \dfrac{K_2[M]}{K_3 V^2}$ $= \dfrac{3D_h K_2 K_3^{1/3}}{2^{5/3} K_1^{2/3} [O^{2-}]^{2/3} [M]^{1/3}} P^{1/3}$ |
| reaction equilibrium State | $M^{p+} + e^- \xleftarrow{\quad} M^{(p-1)+}$ | $M^{p+} + e^- \xrightarrow{\quad} M^{(p-1)+}$ | $M^{p+} \xrightarrow{\quad} M^{(p-1)+} + h^+$ |

[a] Provided $P \gg 4[O^{2-}]^2 K_1^2 / K_3^2 [M]^4$ ; $(D_e / D_V)^4 [O^{2-}]^2 K_1^2 / 4[M]^6$ .

[b] Provided $P \ll 32(D_V / D_h)^3 [M][O^{2-}]^2 K_1^2 / (K_2^3 K_3)$ .

The dependencies summarized in Tables 1 and 2 for the oxygen chemical diffusivity $D^\delta$ are displayed graphically in Fig. 1 for specific values of $K_2$ and $K_3$ , as inset in the figure frame.

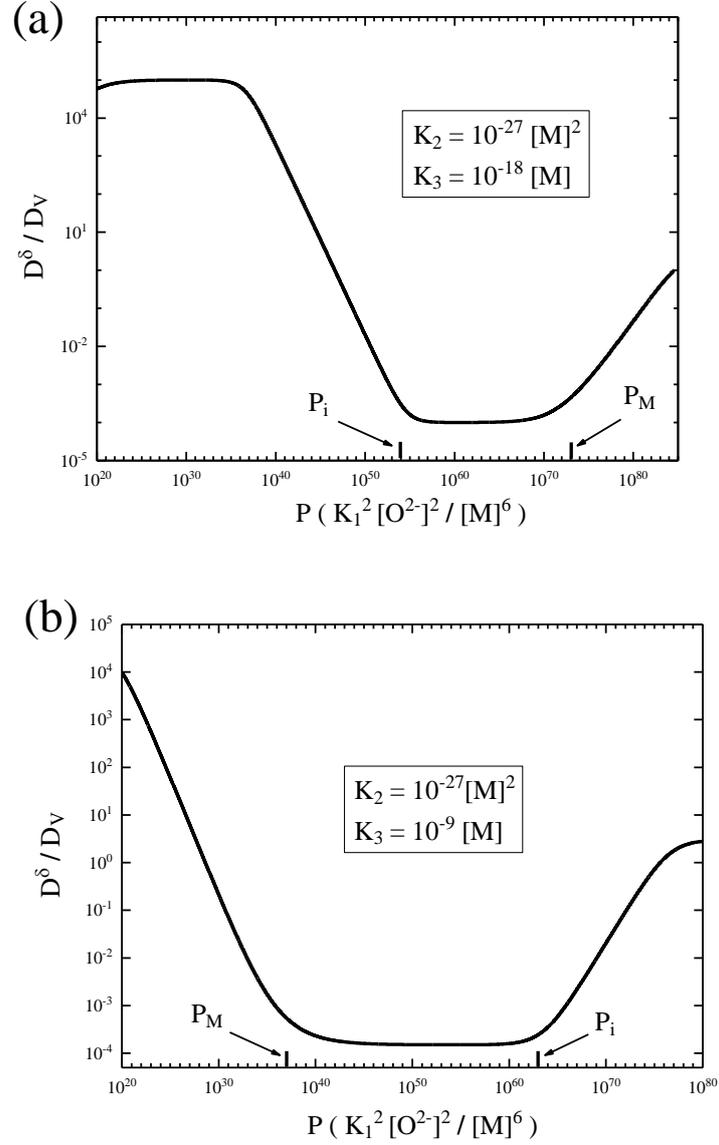

FIG. 1: Ratio of the oxygen chemical diffusivity to the intrinsic vacancy diffusivity for an oxide crystal host of $8\,eV$ energy gap as function of the partial ambient oxygen pressure calculated by Eq. (1). (a) The $M^{p+}$ ion state located $1.0\,eV$ below mid-gap; (b) The $M^{p+}$ state located $1.0\,eV$ above mid-gap. The electron and hole diffusivities $D_e$ and $D_h$, respectively, were taken equal, $10^5$ times larger than the intrinsic vacancy diffusivity $D_V$. The reaction constants $K_2$ and $K_3$ relevant to each case are inset in the figures' frames.

Next we discuss the concentration and reaction rate profiles of various species during diffusion for the different regions defined in Tables 1 and 2 under oxidation or reduction. We limit considerations to a semi-infinite sample.

We consider first the pressure regions in Tables 1 ($P_i << P << P_M$) and 2 ($P_M << P << P_i$), where $D^\delta$ is practically constant. The diffusivities may be readily interpreted as trap-controlled free-hole diffusion per said region in Table 1, and trap-controlled free-electron diffusion per said region in Table 2. The role of traps in the first case is played by the $M^{(p-1)+}$ ions, and in the second case by the $M^{p+}$ ions.

Solution of Eq. (2) for a constant $D^\delta$ is [12]

$$V = V_S + (V_0 - V_S)\,\mathrm{erf}\!\left( x \big/ 2\sqrt{D^\delta \cdot t} \right) \; , \tag{3}$$

where $V_S$ is the surface ($x = 0$) vacancy concentration, and $V_0$ is the initial ($t = 0$), spatially constant vacancy concentration throughout the sample. Eq. (3) is valid for both oxidation ($V_0 > V_S$) and reduction ($V_0 < V_S$). The vacancy diffusion entails a change in the metal ionization state. The rate of concentration change of metal ions at their $(p-1)+$ state (the "reaction rate") is

$$R(x,t) \equiv \partial [M^{(p-1)+}] \big/ \partial t \; . \tag{4}$$

Notably, $[M^{(p-1)+}] \approx 2V$ ; then $R(x,t) \cong 2\,\partial V / \partial t$ .

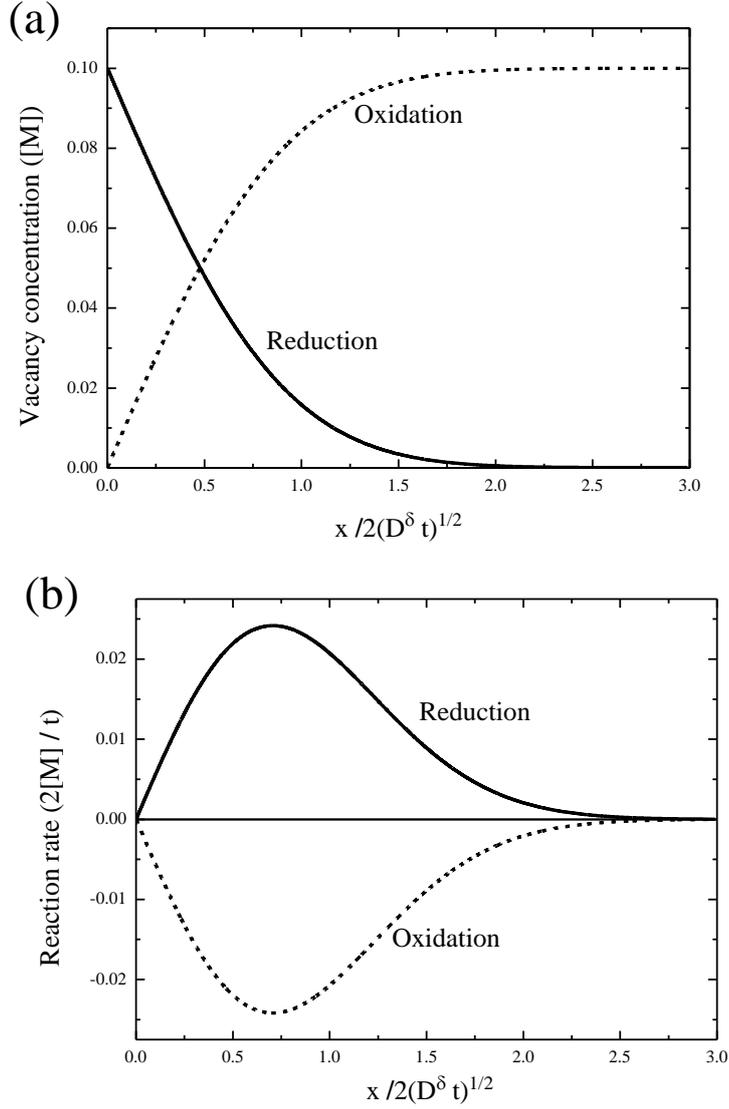

FIG. 2: (a) Vacancy concentration profiles under reduction ($V_S = 10^{-1}[M]$, $V_0 = 10^{-4}[M]$) and oxidation ($V_S = 10^{-4}[M]$, $V_0 = 10^{-1}[M]$) conditions in the region of constant effective diffusivity (Tables 1 and 2). (b) Reaction rate profiles related to same conditions as in (a).

Fig. 2 shows the vacancy concentration and reaction-rate profiles per Eq. (4). The reaction rate is positive for reduction ($V_0 < V_S$), and negative for oxidation ($V_0 > V_S$). It exhibits a broad peak at $x_{peak} = \sqrt{2D^\delta t}$, of width $\sim 1.1\, x_{peak}$. The propagation velocity of the reaction rate into the crystal slows down with time as $t^{-1/2}$, and the reaction rate peak height reduces as $t^{-1}$.

Inserting the specific reaction constants $K_2 = 10^9 \, cm^{-6}$; $K_3 = 1 \, cm^{-3}$ (2nd line, Table 2 in Part I), and further assuming $D_h = 10^5 D_V$, one obtains $D^\delta = D_h (K_2 / [M] K_3) \cong 10^{-4} D_V$ per Table 1. Similarly, inserting the specific reaction constant $K_3 = 10^9 \, cm^{-3}$ (1$^{st}$ line, Table 2 in Part I), and further assuming $D_e = 10^5 D_V$, one obtains

$D^\delta = 1.5 D_e K_3 / [M] \cong 1.5 \times 10^{-4} D_V$ per Table 2. These low values of the oxygen chemical diffusivities relative to the vacancy intrinsic diffusivity $D_V$ occur in spite the fact that the process "driving force" is the vacancy diffusion; also, in spite the fact that the free hole and electron diffusivities were assumed $10^5$ times larger than $D_V$. Apparently, the trap-controlled nature of the free carriers' diffusion dominates over the vacancy diffusion even though the carriers' diffusion constants proper are very large.

In both low-pressure regions $P << P_i$ and $P << P_M$ per Tables 1 and 2, respectively, most dopant metal ions exist in the $M^{(p-1)+}$ state, and the reaction equilibrium state is identical: $M^{p+} + e^- \xleftarrow{\hspace{1cm}} M^{(p-1)+}$. The effective diffusivities are described by the same expression, displaying an $([M] - 2V)^{-2}$ dependence on the vacancy concentration. We denote the surface concentration of $M^{p+}$ ions as $[M^{p+}]_S$. Then, the oxygen chemical diffusivity is

$$D^\delta = D_e \frac{K_3 [M]}{([M] - 2V)^2} = \frac{D^*}{([M^{p+}]/[M^{p+}]_S)^2} \quad , \qquad (5)$$

where $D^* = D_e K_3 [M] / ([M^{p+}]_S)^2$.

We invoke Seeger's parametric solution [9]: One defines a constant $A$ as the solution of the equation

$$1 - A\sqrt{\pi} \, \mathbf{exp}(A^2) \, \mathbf{erfc}(A) = \frac{[M^{p+}]_S}{[M^{p+}]_0} \quad , \tag{6}$$

where $[M^{p+}]_0$ is the $M^{p+}$ ion concentration at $t = 0$. For a free parameter $\lambda \geq A$

$$y \equiv \frac{x}{2\sqrt{D^* t}} = \left\{ \lambda - A \, \mathbf{exp}(A^2 - \lambda^2) - \lambda A \pi^{1/2} \, \mathbf{exp}(A^2) \big[ \mathbf{erfc}(A) - \mathbf{erfc}(\lambda) \big] \right\} , \tag{7a}$$

and

$$[M^{p+}] \big/ [M^{p+}]_S = \frac{1}{1 - A\sqrt{\pi} \, \mathbf{exp}(A^2) \, \mathbf{erfc}(A) \left[ 1 - \dfrac{\mathbf{erfc}(\lambda)}{\mathbf{erfc}(A)} \right]} \quad , \tag{7b}$$

For demonstration of the above solution under reduction and oxidation we use $K_3 = 1 \, \text{cm}^{-3}$, and the initial and boundary conditions per Table 3 below. The latter were selected to cover the entire relevant pressure ranges.

TABLE 3: Initial and boundary conditions used in Figs. 3 and 4, both with $K_3 = 1 \, \text{cm}^{-3}$ and $[M] = 10^{18} \, \text{cm}^{-3}$

|  | Initial condition | Boundary condition |
|---|---|---|
| Reduction | $[M^{p+}]_0 = 10^{13.5} \, \text{cm}^{-3}$ | $[M^{p+}]_S = 10^{9.5} \, \text{cm}^{-3}$ |
| Oxidation | $[M^{p+}]_0 = 10^{9.5} \, \text{cm}^{-3}$ | $[M^{p+}]_S = 10^{13.5} \, \text{cm}^{-3}$ |

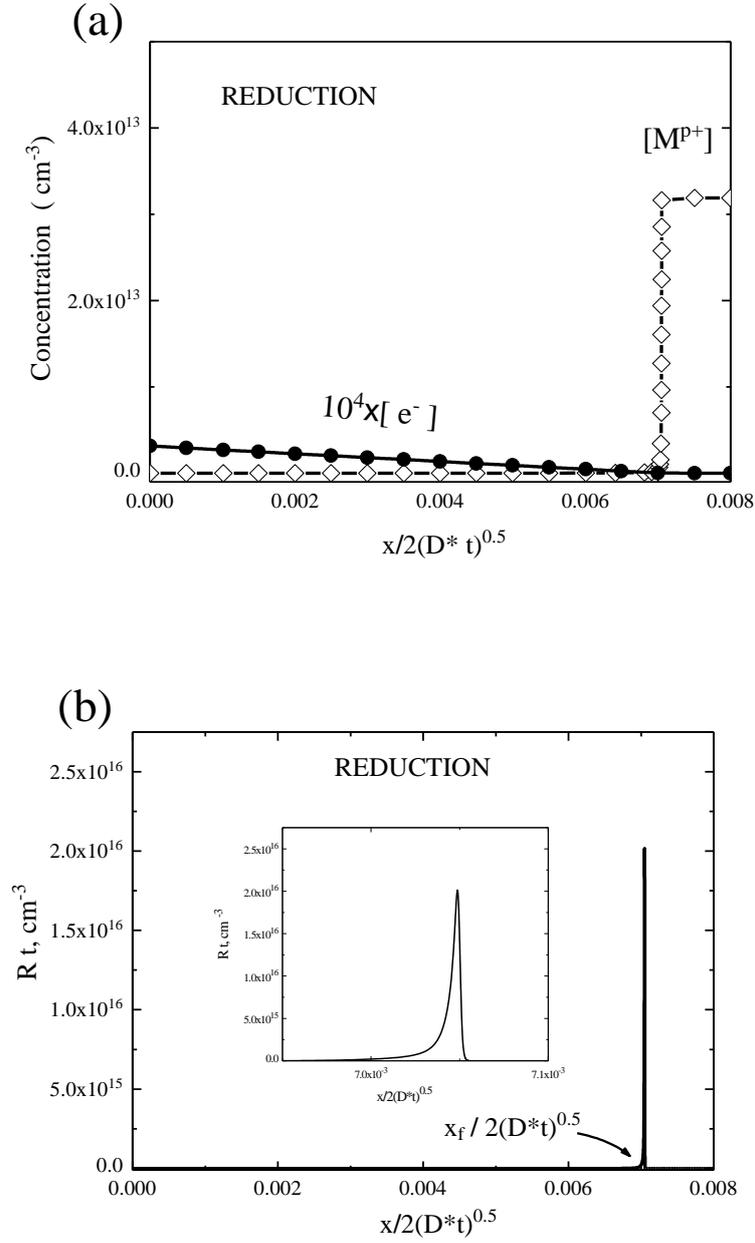

FIG.3: (a) $M^{p+}$ ion and free electron concentrations per Eq. (7) as functions of depth below the sample surface during reduction for $K_3 = 1 \, cm^{-3}$, and for initial and boundary conditions $[M^{p+}]_0 = 10^{13.5} \, cm^{-3}$, and $[M^{p+}]_S = 10^{9.5} \, cm^{-3}$; (b) Reaction rate-time product as function of reduced depth below the surface. Inset is a blow-up of the position scale in the vicinity of the reaction rate peak (reaction front).

In Fig. 3(a), the $M^{p+}$ ion and free electron concentrations are plotted as functions of depth below the sample surface during reduction. The $M^{p+}$ ion concentration is virtually zero close to the surface (namely for $x < x_f = 7 \times 10^{-3} \times 2\sqrt{D^* t}$ ), but rises sharply at $x_f$ towards the initial concentration. Notably, all $M^{p+}$ concentrations are very small compared to $[M^{(p-1)+}] \approx [M]$. In Fig. 2(b) the reaction rate is plotted as function of depth below the surface. A peak in the reaction rate is obtained at $x \cong x_f$, establishing an unequivocal reaction front. The inset in the figure is just a blow-up of the position scale in the vicinity of the reaction front. Importantly, a reaction front could occur only when $[e^-]_S / K_3 \gg 1$, where $[e^-]_S$ is the surface free electron concentration, a condition satisfied for Fig. 3(b). Recall that the reaction involved is $M^{p+} + e^- \xleftarrow{\hspace{1cm}} M^{(p-1)+}$ (Tables 1 and 2).

Fig. 3 gives $x_f = \sqrt{2 \times 10^{-4} D^* t} = \sqrt{2 \times 10^{-5} D_e t}$. Recalling that $D_e \approx 10^5 D_V$ one obtains $x_f \cong \sqrt{2 D_V t}$. Thus the apparent diffusivity dominating the reaction front position is in fact the oxygen vacancy diffusivity $D_V$; this in-spite the fact that the oxygen chemical diffusivity $D^\delta$ is dominated by the electron diffusivity $D_e$ assumed $10^5$ times larger than $D_V$. The apparent effect of the large oxygen chemical diffusivity is to render the $M^{p+}$ concentration behind the reaction front small, and virtually constant.

In fact, an expression for $x_f$ may be obtained by a simple model that equates the free electron current density to the product of $M^{p+}$ ion density with the progressing speed of the reaction front $dx_f / dt$ (Ref. 13, Eqs. (16)-(18)). The expression obtained is $x_f^2 = 2D_e t([e^-]_S / [M^{p+}]_0)$, where $[M^{p+}]_0$ is the initial $M^{p+}$ ion concentration. Inserting the latter numerical values yields the same estimate for $x_f$ as in Fig. 3.

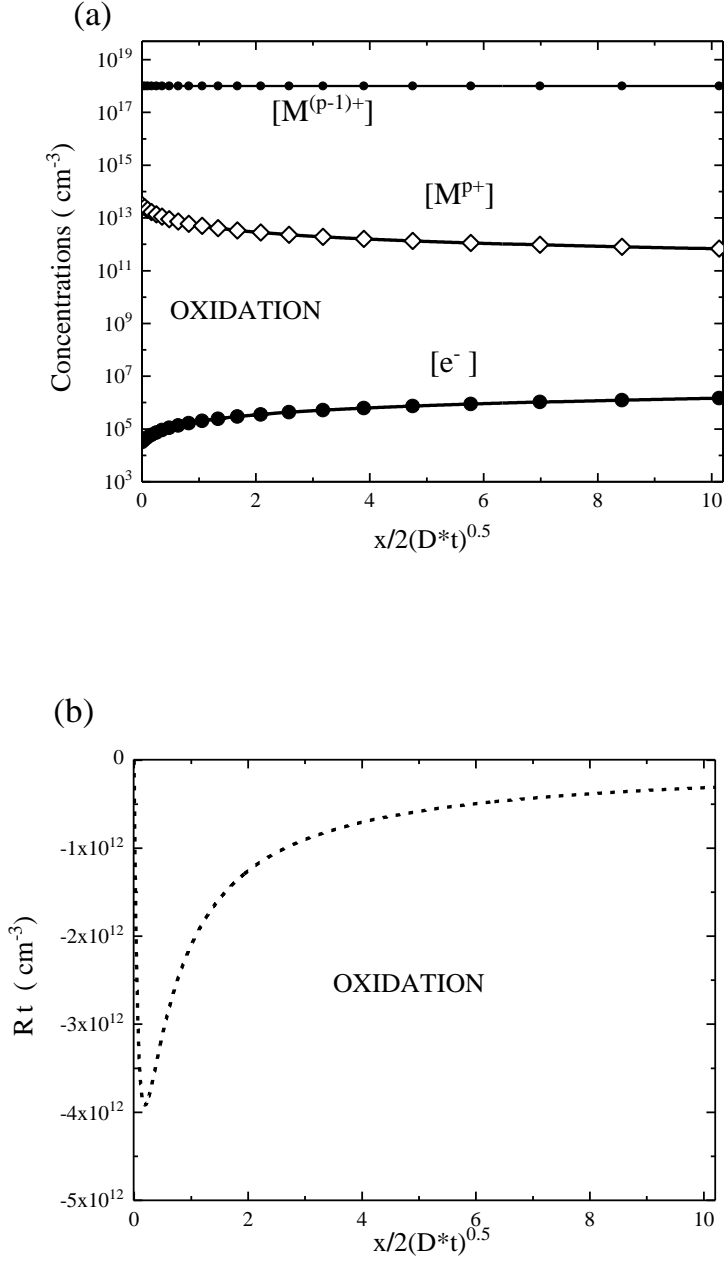

FIG. 4: (a) $M^{p+}$ and $M^{(p-1)+}$ ions, and free electron $e^-$ concentrations per Eq. (7) as functions of depth below the sample surface during oxidation for $K_3 = 1\,\mathrm{cm}^{-3}$, and for initial and boundary conditions $[M^{p+}]_0 = 10^{9.5}\,\mathrm{cm}^{-3}$, and $[M^{p+}]_s = 10^{13.5}\,\mathrm{cm}^{-3}$; (b) Reaction rate–time product as function of reduced depth below the surface.

In Fig. 4(a), the $M^{p+}$ ion and free electron concentrations are plotted as functions of depth below the sample surface during oxidation. The $M^{p+}$ ion concentration increases near the surface, exhibiting a monotonic decrease towards the initial value $[M^{p+}]_0$ deep below the surface. The free electron concentration exhibits a corresponding decrease at the surface. Note that the said $M^{p+}$ ion concentration is orders of magnitude smaller than the $M^{(p-1)+}$ ion concentration. Thus, changes in the latter are not visible on the logarithmic scale used. Obviously, $[M^{(p-1)+}]$ reduces at the surface, as the sum $[M^{(p-1)+}]+[M^{p+}]$ is constant (Eq. (8) in Part I). In Fig. 4(b) the reaction rate is plotted as function of depth below the surface. The reaction rate is negative, as expected for an oxidation case (Fig. 2(b)). A peak in the absolute value of the reaction rate is obtained at $x \cong 0.2 \times 2\sqrt{D^* t}$; the peak is, however, very broad (approximately $2\sqrt{D^* t}$), also exhibiting an extended tail reaching into the crystal. Thus no reaction front may be defined. In some respects, the diffusion characteristics are similar to those obtained for a constant diffusivity (Fig. 2). Still, while for the constant diffusivity case displayed in Fig. 2, the apparent diffusivity was of the order of $10^{-4} D_V$, in the present case the apparent diffusivity for the reaction-rate peak position is quite smaller, only of the order of $10^{-6} - 10^{-5} D_V$.

In both high-pressure regions, $\tilde{P} \gg \tilde{P}_M$ and $\tilde{P} \gg \tilde{P}_I$ in Tables 1 and 2, respectively, most dopant metal ions exist in the $M^{p+}$ state, and the reaction equilibrium state is identical: 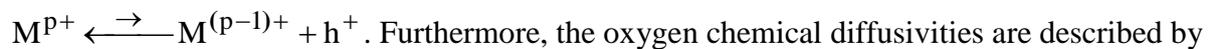 $M^{p+} \xleftarrow{\quad\longrightarrow\quad} M^{(p-1)+} + h^+$. Furthermore, the oxygen chemical diffusivities are described by the same expression, displaying a $V^{-2}$ dependence on the oxygen vacancy concentration.

We denote the surface concentration of $M^{(p-1)+}$ ions as $[M^{(p-1)+}]_S$. Then, the oxygen chemical diffusivity may be written as

$$D^\delta = D_h \frac{3}{8} \frac{K_2}{K_3} \frac{[M]}{V^2} = \frac{\hat{D}}{[M^{(p-1)+}]^2 \big/ [M^{(p-1)+}]_S^2} \qquad , \qquad (8)$$

where $\hat{D} = 1.5 D_h (K_2 / [M] K_3)([M] \big/ [M^{(p-1)+}]_S)^2$.

In the present case one may readily use Eqs. (6) - (7) as a solution of the diffusion equation [9] by simply replacing $D^*$ with $\hat{D}$ in (7a), and $[M^{p+}]$ and $[M^{p+}]_S$ by $[M^{(p-1)+}]$ and $[M^{(p-1)+}]_S$, respectively, in (7b) and (6), where $[M^{(p-1)+}]_0$ is the $M^{(p-1)+}$ ion concentration at $t = 0$. Thus technically, solutions of these different cases are identical for properly symmetric cases. These symmetric cases are specified in Table 4. The gross feature indicating each case is the occurrence or non-occurrence of a reaction front.

TABLE 4: Summary of reaction front occurrence under oxidation or reduction for the different extreme pressure ranges.

| Pressure Range; Reaction equilibrium State | $D^\delta$ | Reaction Front Occurrence | |
|---|---|---|---|
| | | Oxidation | Reduction |
| $P \ll P_1, P_M$ ;  $M^{p+} + e^- \xleftarrow{\quad} M^{(p-1)+}$ | $D_e \dfrac{K_3[M]}{([M]-2V)^2}$ | No | Yes |
| $P \gg P_1, P_M$ ;  $M^{p+} \xleftarrow{\quad\rightarrow\quad} M^{(p-1)+} + h^+$ | $D_h \dfrac{3}{8} \dfrac{K_2}{K_3} \dfrac{[M]}{V^2}$ | Yes | No |

**V. CONCLUSIONS**

Our present theoretical work proposes that carrying out of experimental measurements of metal-ion doped oxide crystals under dynamic conditions (namely during diffusion) would be useful for assessing the relevant reaction constants. Physically clear solutions of the oxygen vacancy diffusion are obtained in regions defined by $P_i$ and $P_M$. These observations should be very helpful for an experimental work: provided the initial and final oxygen partial pressures are selected to fall within the well defined regions, utilization of the relevant expressions and effects is straightforward, calling for a concentration profile study of one or more of the participating species (electrons, holes, oxygen vacancies, or metal dopant ionic states). A more complex behavior would occur if the initial and final states belong to different regions (for example, Fig. 4 in Ref. 14). In fact, the actual behavior may be represented as a combination among the ones related to the separate regions. This complex behavior issue will be a subject in a future publication.

**Figure captions**

FIG. 1: Ratio of the oxygen chemical diffusivity to the intrinsic vacancy diffusivity for an oxide crystal host of $8\,eV$ energy gap as function of the partial ambient oxygen pressure calculated by Eq. (1). (a) The $M^{p+}$ ion state located $1.0\,eV$ below mid-gap; (b) The $M^{p+}$ state located $1.0\,eV$ above mid-gap. The electron and hole diffusivities $D_e$ and $D_h$, respectively, were taken equal, $10^5$ times larger than the intrinsic vacancy diffusivity $D_V$. The reaction constants $K_2$ and $K_3$ relevant to each case are inset in the figures' frames.

FIG. 2: (a) Vacancy concentration profiles under reduction ( $V_S = 10^{-1}[M]$, $V_0 = 10^{-4}[M]$ ) and oxidation ( $V_S = 10^{-4}[M]$, $V_0 = 10^{-1}[M]$ ) conditions in the region of constant effective diffusivity (Tables 1 and 2). (b) Reaction rate profiles related to same conditions as in (a).

FIG.3: (a) $M^{p+}$ ion and free electron concentrations per Eq. (7) as functions of depth below the sample surface during reduction for $K_3 = 1\,cm^{-3}$, and for initial and boundary conditions $[M^{p+}]_0 = 10^{13.5}\,cm^{-3}$, and $[M^{p+}]_S = 10^{9.5}\,cm^{-3}$; (b) Reaction rate-time product as function of reduced depth below the surface. Inset is a blow-up of the position scale in the vicinity of the reaction rate peak (reaction front).

FIG. 4: (a) $M^{p+}$ and $M^{(p-1)+}$ ions, and free electron $e^-$ concentrations per Eq. (7) as functions of depth below the sample surface during oxidation for $K_3 = 1\,cm^{-3}$, and for initial and boundary conditions $[M^{p+}]_0 = 10^{9.5}\,cm^{-3}$, and $[M^{p+}]_s = 10^{13.5}\,cm^{-3}$; (b) Reaction rate–time product as function of reduced depth below the surface.